\documentclass[final,times,twocolumn,sort&compress]{elsarticle}
\usepackage{caption}
\usepackage{subfigure,dcolumn}
\usepackage{amsmath}
\usepackage{epstopdf}
\usepackage{mhchem}
\usepackage{upgreek}
\usepackage{bm}
\usepackage{color}
\usepackage{graphicx}
\usepackage{lmodern}
\usepackage{mathtools}
\usepackage{multirow}
\usepackage{makecell}
\usepackage{textcomp}
\usepackage{ulem}
\usepackage{xurl}
\usepackage{xspace}

\newcommand{\snn} {\ensuremath{\sqrt{s_{\rm NN}}}\xspace}
\newcommand{\pt} {\ensuremath{p_{\rm T}}\xspace}
\newcommand{\vtwo} {\ensuremath{v_2}\xspace}
\newcommand{\lp} {\ensuremath{\Lambda-p}\xspace}

\newcommand{\dd} {\ensuremath{\Delta\delta}\xspace}
\newcommand{\dg} {\ensuremath{\Delta\gamma}\xspace}

\begin{document}

\title{Assessing background effects in search of the chiral vortical effect in relativistic heavy-ion collisions}

\author[label1,label2]{Chunzheng Wang}
\author[label1,label2]{Jie Wan}
\author[label3]{Jinfeng Liao\corref{cor1}}
\ead{liaoji@iu.edu}
\author[label1,label2]{Yugang Ma\corref{cor1}}
\ead{mayugang@fudan.edu.cn}
\author[label4]{Shuzhe Shi\corref{cor1}}
\ead{shuzhe-shi@tsinghua.edu.cn}
\author[label1,label2]{Qiye Shou\corref{cor1}}
\ead{shouqiye@fudan.edu.cn}
\author[label1,label2]{Zhengqing Wang}
\author[label1,label2]{Kegang Xiong}
\author[label1,label2]{Song Zhang}
\author[label5,label2]{Liang Zheng}
\cortext[cor1]{Corresponding authors}
\address[label1]{Key Laboratory of Nuclear Physics and Ion-beam Application (MOE), Institute of Modern Physics, Fudan University, Shanghai 200433, China}
\address[label2]{Shanghai Research Center for Theoretical Nuclear Physics, NSFC and Fudan University, Shanghai 200438, China}
\address[label3]{Physics Department and Center for Exploration of Energy and Matter, Indiana University, Indiana 47408, USA}
\address[label4]{Department of Physics, Tsinghua University, Beijing 100084, China}
\address[label5]{School of Mathematics and Physics, China University of Geosciences (Wuhan), Wuhan 430074, China}

\begin{abstract}
The search for the Chiral Vortical Effect (CVE) in relativistic heavy-ion collisions is carried out by measuring azimuthal correlators for baryon pairs such as $\Lambda$ and protons. Experimental results from the ALICE collaboration show significant separations in these observables, however, the interpretation remains unclear. It is believed that background contributions from baryon production mechanisms may play an important role.
Using three phenomenological models, the Blast Wave, AMPT, and AVFD+UrQMD, we systematically investigate the background effects in Pb--Pb collisions at \snn = 5.02 TeV. We demonstrate that local baryon conservation, as well as hadronic annihilation processes, can significantly influence the correlators. The feed-down contribution from secondary protons is also estimated. Our study provides a foundation for disentangling background mechanisms and further facilitates the search for the CVE.
\end{abstract}

\maketitle

%%%%%%%%%%%%%%%%%%%%%%%%%%%%%%%%%%%%%%%%%%%%%%%%%%%%%%%%%
\section{Introduction} \label{sec:intro}

In relativistic heavy-ion collisions, it has been proposed that the interplay between the local chiral anomaly and the intense magnetic and vortical fields produced in non-central collisions can lead to various anomalous chiral phenomena~\cite{Kharzeev:2008, Kharzeev:2016}. Investigating these effects may offer insights into the topological structure of vacuum gauge fields and uncover potential local $\cal P$ (parity) and/or $\cal CP$ (charge-parity) violations in strong interactions~\cite{Kharzeev:2021}. The chiral magnetic effect (CME) in the quark-gluon plasma (QGP) is one of the most prominent and has been thoroughly studied both theoretically and experimentally~\cite{Zhao:2019, Li:2020}. 

In addition to the CME, the chiral vortical effect (CVE)~\cite{Kharzeev:2011, Jiang:2015} is another key phenomenon. In this case, the vorticity field $\omega$ combined with a baryon chemical potential $\mu_{\rm B}$ generates an effective magnetic field. Such vorticity emerges naturally due to the conservation of initial angular momentum in non-central heavy-ion collisions. The role and quantitative estimates of vorticity have been extensively discussed in studies of both global and local polarization phenomena~\cite{STAR:2017, Becattini:2020}.

The CVE is predicted to produce a baryonic dipole moment with respect to the reaction plane. By measuring the azimuthal emission of final-state baryons, it is possible to detect the CVE-induced baryonic charge separation through the $\gamma$ and $\delta$ correlators~\cite{Voloshin:2004, Kharzeev:2006}:
$\gamma \equiv {\rm cos}(\phi_\alpha+\phi_\beta-2\Psi), \quad \delta \equiv {\rm cos}(\phi_\alpha-\phi_\beta)$,
where $\phi_\alpha$ and $\phi_\beta$ are the azimuthal angles of two particles of interest (baryons), and $\Psi$ is the azimuthal angle of the reaction plane. These measurements are typically performed with same-sign (SS) and opposite-sign (OS) charge combinations, and the differences between these combinations are analyzed to explore potential signals:
$\Delta\gamma \equiv \gamma^{\rm OS}-\gamma^{\rm SS}, \quad \Delta\delta \equiv \delta^{\rm OS}-\delta^{\rm SS}$.

The ALICE~\cite{Shou:2024, Acharya:2024} and STAR collaborations~\cite{Chen:2024} have conducted the CVE measurements across various collision energies and systems. In particular, $\Lambda$-proton ($p$) pairs are chosen as the probe, since the charge neutrality of the $\Lambda$ helps minimize influences from electric charge-dependent effects such as CME~\cite{Kharzeev:2011, Jiang:2015}.
ALICE measurements in Pb--Pb collisions at \snn = 5.02 TeV~\cite{Wang:2024} show a notable separation in both $\gamma$ and $\delta$ correlators, approximately an order of magnitude larger than that observed in charged hadron ($h-h$) correlations, while $\Lambda-h$ correlations are nearly negligible. Similarly, STAR data from Au--Au collisions at RHIC BES energies~\cite{Xu:2024} also indicate stronger \lp correlations compared to $h-h$ pairs. Additionally, ALICE data demonstrate a clear kinematic hierarchy across different transverse momentum (\pt) and pseudorapidity ($\eta$) intervals.

These observations raise the question of whether the measured results are consistent with CVE expectations.
In contrast to the study of the CME, which has benefitted from over a decade of background characterization and the development of multiple methods to extract signal fractions and upper limits, the interpretation of CVE measurements remains unclear. While Ref.~\cite{Frenklakh:2024} suggests that ALICE data can be explained within a hydrodynamic framework when a realistic chiral chemical potential is incorporated, few studies have yet systematically examined background sources in these observables.

Since the CVE measurements involve correlations between particles carrying baryonic charge, it is reasonable to infer that potential backgrounds are linked to baryon production intertwined with azimuthal emission. Analogous to the role of local charge conservation (LCC) in the CME study~\cite{Bzdak:2011}, the effects of global and/or local baryon conservation (LBC) should be assessed.
To this end, we employ three widely used phenomenological models: the blast wave model (BW), a multi-phase transport model (AMPT), and event-by-event anomalous viscous fluid dynamics (AVFD) coupled with UrQMD, to examine background contributions from primordial and secondary baryons.
In the following sections, the model and analysis method are presented in Sec.\ref{method}, followed by the main results and discussions in Sec.\ref{results}, and a brief summary in Sec.\ref{sum}.

%%%%%%%%%%%%%%%%%%%%%%%%%%%%%%%%%%%%%%%%%%%%%%%%%%%%%%%%%
\section{Model and analysis method} \label{method}

The BW model~\cite{Retiere:2004, Tomasik:2009} generates an expanding, locally thermalized fireball that decays into fragments and subsequently emits hadrons. The hadrons are assumed to be in thermal equilibrium according to a Boltzmann distribution characterized by a kinetic freeze-out temperature. The phase space distribution is modeled under the assumption that the radial expansion velocity increases linearly with the distance from the system center. 
In this study, only two baryon species of interest, $\Lambda$ and $p$, are generated for clarity and focus.
To introduce the local baryon conservation, $\Lambda$ and $p$ are emitted in pairs from some spatial points uniformly distributed within an ellipse~\cite{Wu:2023}, with each pair satisfying total baryon number neutrality, i.e., $\Lambda-\bar{p}$ or $\bar{\Lambda}-p$.
The momenta of particles within a pair are sampled independently and then boosted collectively to follow a common flow velocity. 
For the rest of the spatial points, only a single $\Lambda$ or $p$ is emitted with random baryonic charge. The fraction of pair-emitting points, $f_{\rm LBC}$, is used to control the strength of LBC.
Since the experimental balance function~\cite{Bass:2000} between $\Lambda$ and $p$ is currently unmeasured, we tune $f_{\rm LBC}$ to reproduce the ALICE CVE results, as will be presented in the following section.

The hybrid transport model AMPT~\cite{Lin:2005, Lin:2021} is known for effectively describing the production and collectivity of the final state hadrons. The string melting version comprises four subroutines simulating successive stages of the collision: HIJING for initial parton condition~\cite{Gyulassy:1994}, ZPC for partonic evolution~\cite{Zhang:1998}, a simple quark coalescence for hadronization~\cite{Fries:2008}, and ART for hadronic rescatterings and interactions~\cite{Li:1995}. Here we mainly focus on baryon correlations arising from coalescence, so ART module is disabled, and the investigation of hadronic effects is performed in AVFD+UrQMD.

The AVFD framework~\cite{Shi:2018, Jiang:2018} simulates the heavy-ion collision with hydrodynamic evolution and chiral transport, followed by Cooper--Frye hadronization that incorporates LCC/LBC at a freeze-out temperature $T_{\mathrm{fo}} = 160~\mathrm{MeV}$. After that, the UrQMD model~\cite{Bass:1998, Bleicher:1999} is used to simulate hadronic scatterings and resonance decays. 
Since this study focuses on backgrounds, i.e., LBC and resonance decays, chiral anomalous effects are therefore disabled. 
The implementation of the LBC in AVFD is similar to that used in the BW: a randomly selected fraction of particles, denoted as $P_\mathrm{LBC}$, are produced in pairs alongside their corresponding antiparticles within the same freeze-out cell. 
Two configurations are considered: $P_\mathrm{LBC} = 0$ (LBC disabled) and $P_\mathrm{LBC}=30\%$ ($30\%$ of particles created in pairs). 
Additionally, the impact of hadronic rescatterings and decays is evaluated by comparing correlator values before and after the UrQMD hadron cascade.

All three models introduced above are simulated for Pb--Pb collisions at \snn = 5.02 TeV. In accordance with established parameterizations, each model is tuned to reproduce the fundamental hadronic spectra and anisotropic flow at LHC energies~\cite{Adam:2016, Acharya:2018}, ensuring consistency with experimental data.

%%%%%%%%%%%%%%%%%%%%%%%%%%%%%%%%%%%%%%%%%%%%%%%%%%%%%%%%%
\section{Results and discussions} \label{results}

\subsection{Centrality and kinematic dependences of \dd and \dg}

\begin{figure}[!htb]
\includegraphics[width=\linewidth]{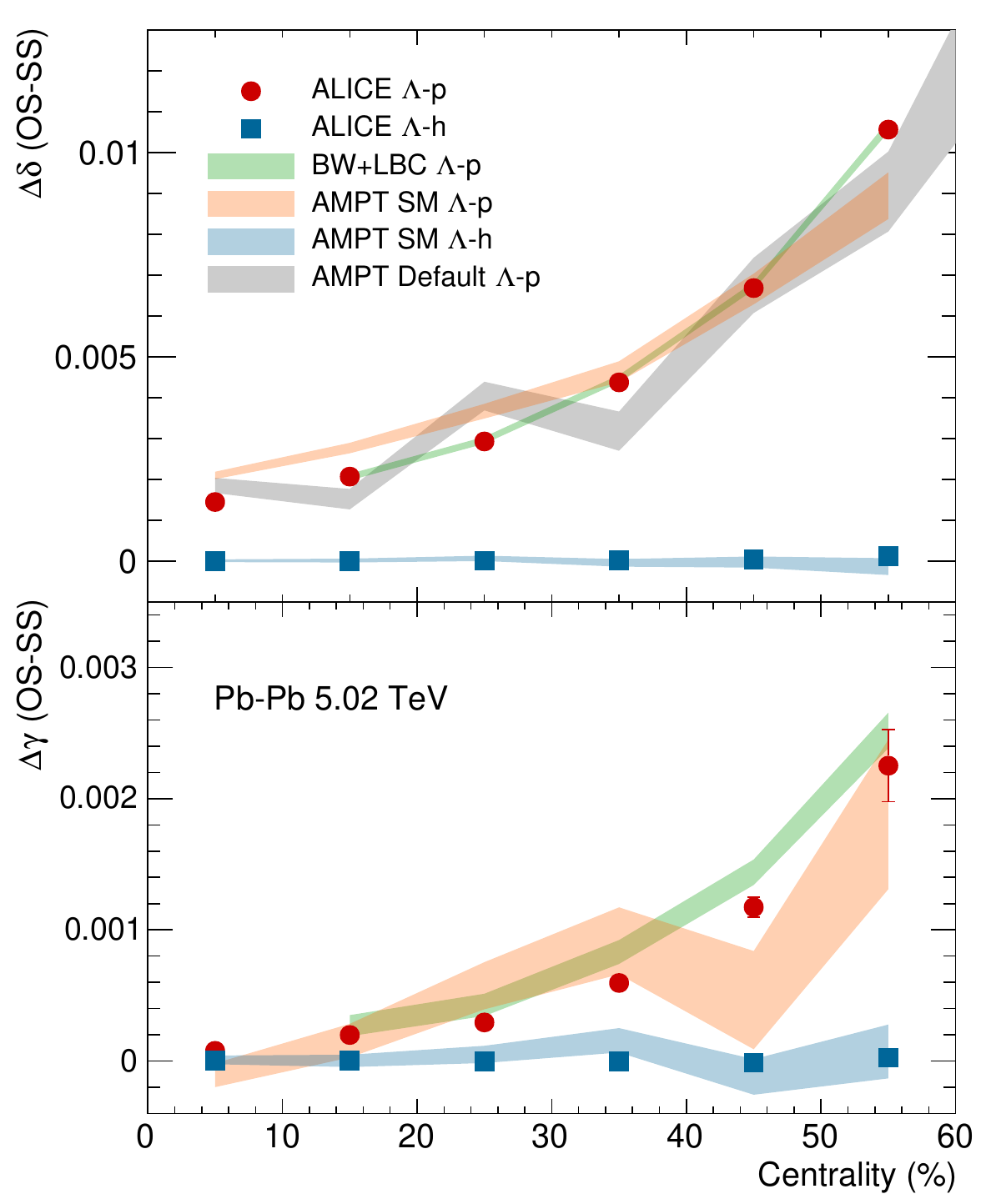}
\caption{(Color online) Centrality dependence of \dd and \dg correlators for \lp and $\Lambda-h$ pairs in Pb--Pb collisions at 5.02 TeV. Calculations from the BW and AMPT are compared with ALICE data~\cite{Wang:2024}.}
\label{fig:fig_cent}
\end{figure}

Figure~\ref{fig:fig_cent} shows the centrality dependence of \dd and \dg in Pb--Pb collisions at 5.02 TeV, obtained with the BW and AMPT models and compared with ALICE data. Both observables in two models increase toward peripheral collisions and reproduce the correct magnitude. In BW, as aforementioned, the parameters are tuned to match ALICE data, so the close agreement is expected. The $f_{\rm LBC}$ rises from $\approx$3.4\% in 10–20\% centrality to $\approx$ 5\% in 50–60\%, opposite to the $f_{\rm LCC}$ behaviour in our previous study~\cite{Wu:2023}, which peaked in central collisions. This suggests possible different mechanisms for baryon-number and charge conservation. 
BW further reveals that \dd is governed mainly by the radial-flow velocity $\beta_{\rm T}$, whereas \dg is controlled by the elliptic-flow coefficient $\rho_2$. In fact, \dg scales linearly with the \vtwo of the event, the proton, and the $\Lambda$. From such a straightforward scenario, it is therefore expected that the CVE signal could also be extracted with event-shape engineering (ESE)~\cite{Schukraft:2013}, following the same strategy already employed for the CME~\cite{Acharya:2018}.

In addition to the BW, the AMPT results are noteworthy: without designed parameter tuning, AMPT simultaneously reproduces the ALICE \dd and \dg data across all centralities within uncertainties. Tracing the full evolution process in AMPT, the agreement is found to be attributed to two baryon-production mechanisms automatically embedded:
(1) Intrinsic baryon conservation inherited from HIJING -- the global (in a whole event) and local (from the same string breakup) baryon anti-baryon conservation;
(2) Partonic coalescence -- when an (anti-)baryon is formed, the local (anti-)quark density drops, suppressing further (anti-)baryon production nearby.
These mechanisms can be easily tested by disabling coalescence through the default AMPT rather than the string melting version, which leaves \dd essentially unchanged but breaks the description of \dg. Since the default AMPT yields an incorrect \vtwo, its failure on reproducing \dg also aligns with the conclusion from the BW that \dg strongly depends on \vtwo.
Note that $\delta$ and $\gamma$ correlators are also connected to two-particle azimuthal correlations, and the role of coalescence in such correlations has been discussed in Refs~\cite{Zhang:2018, Zhang:2022}. A detailed analysis with space-time evolution will be presented in our further work, and for now we simply stress that the experimentally measured correlations already contain, at least partially, the physics of baryon production, regardless of a potential CVE signal.

\begin{figure}[!htb]
\includegraphics[width=\linewidth]{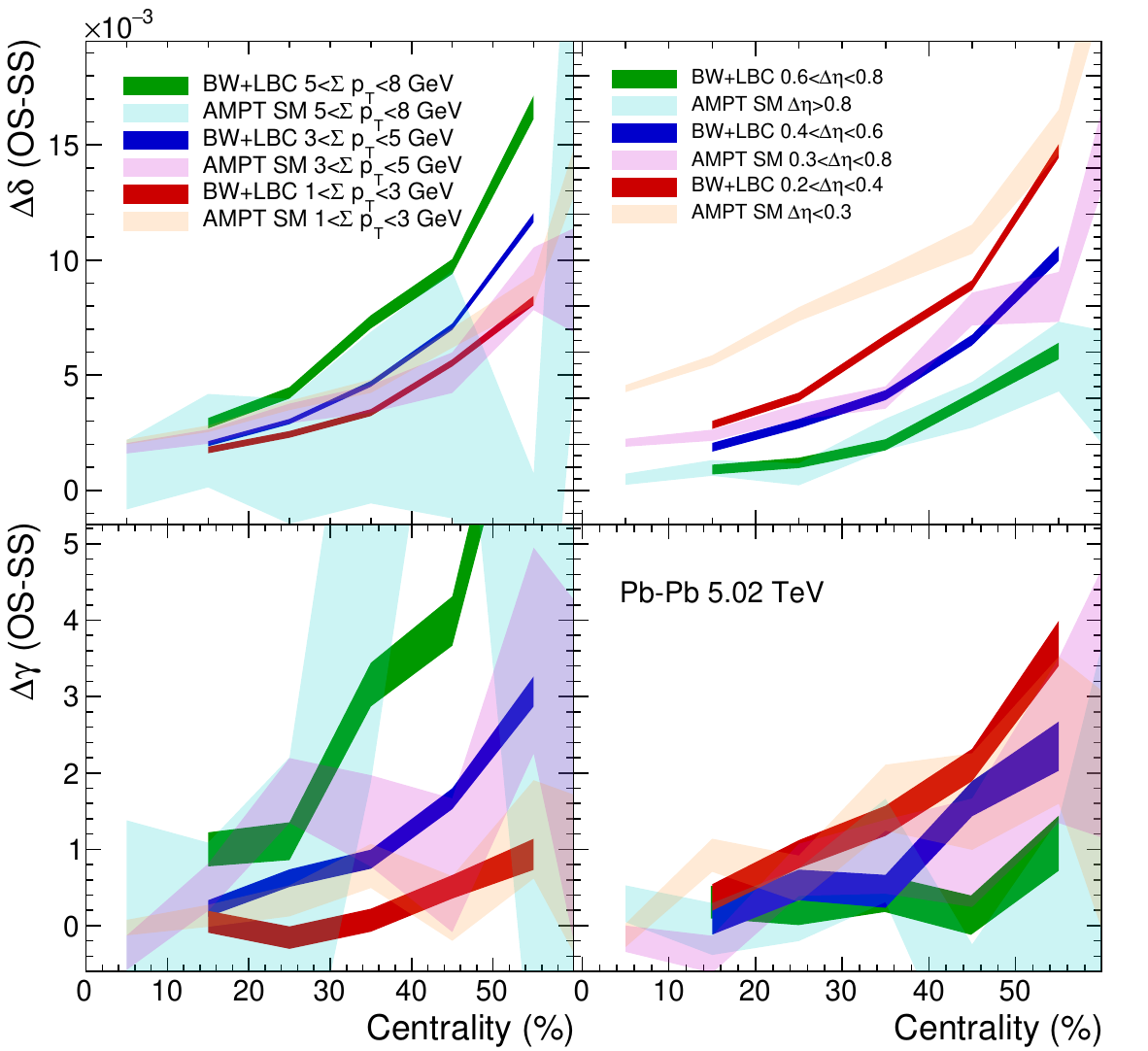}
\caption{(Color online) Centrality dependence of \dd and \dg correlators for \lp pairs in different $\Sigma\pt$ and $\Delta\eta$ kinematic windows. The BW+LBC can reproduce both the measured $\Sigma\pt$ and $\Delta\eta$ hierarchies, while AMPT only describes the $\Delta\eta$ dependence and fails for $\Sigma\pt$ trends.}
\label{fig:fig_pt_eta}
\end{figure}

Examining how observables vary across different kinematic windows is crucial to further probe the underlying mechanisms. Experimentally, the \dd and \dg observables are measured under two selection criteria: 1) the sum of transverse momenta ($\Sigma\pt$) of the $\Lambda$ and $p$, and 2) the pseudorapidity difference ($\Delta\eta$) between the $\Lambda$ and $p$. Both \dd and \dg increase with increasing $\Sigma\pt$ and with decreasing $\Delta\eta$, indicating that the observables are influenced by soft/hard processes and short/long-range effects.
These trends are well described by the BW, as shown in Fig.~\ref{fig:fig_pt_eta}. The $\Sigma\pt$ dependence arises naturally from the BW boost mechanism: higher $\Sigma\pt$ values correspond to stronger collective boost. 
The $\Delta\eta$ dependence can be attributed to the fact that \lp pairs originating from the same spatial points have smaller $\Delta\eta$ than those from different points. Thus, imposing a narrow $\Delta\eta$ selection enhances the contribution from intrinsic \lp correlations.
On the other hand, while the AMPT can reproduce the $\Delta\eta$ trends, it fails to describe the $\Sigma\pt$ dependence. Given the complexity of the model, detailed explanations require further investigation. Overall, comparisons between models and data highlight the existence and significance of background effects from baryon production convolved with collectivity.

%%%%%%%%%%%%%%%%%%%%%%%%%%%%%%%%%%%%%%%%%%%%%%%%%%%%%%%%%
\subsection{Hadronic effects}

\begin{figure}[!hbtp]
\includegraphics[width=0.9\linewidth]{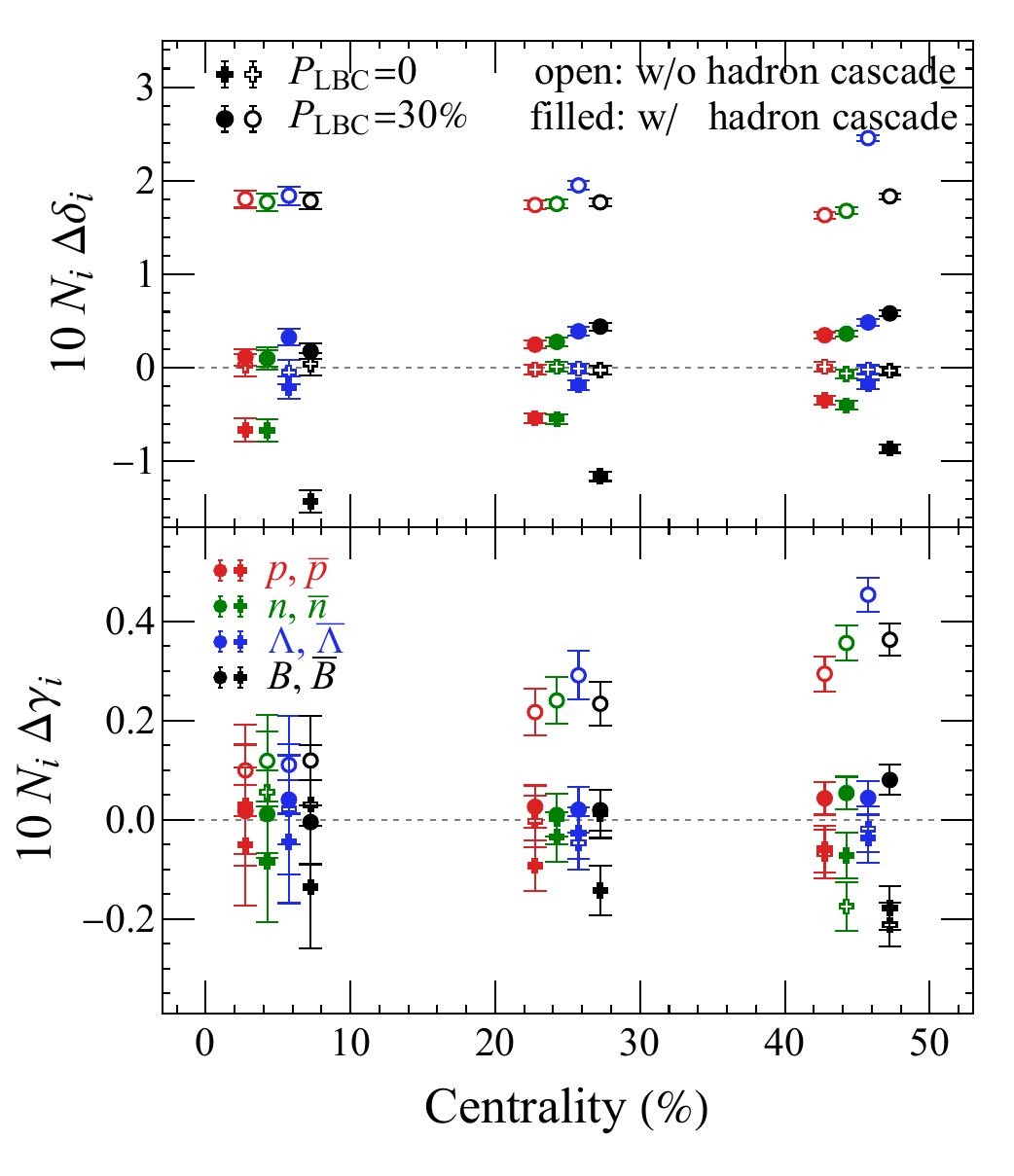}
\includegraphics[width=0.9\linewidth]{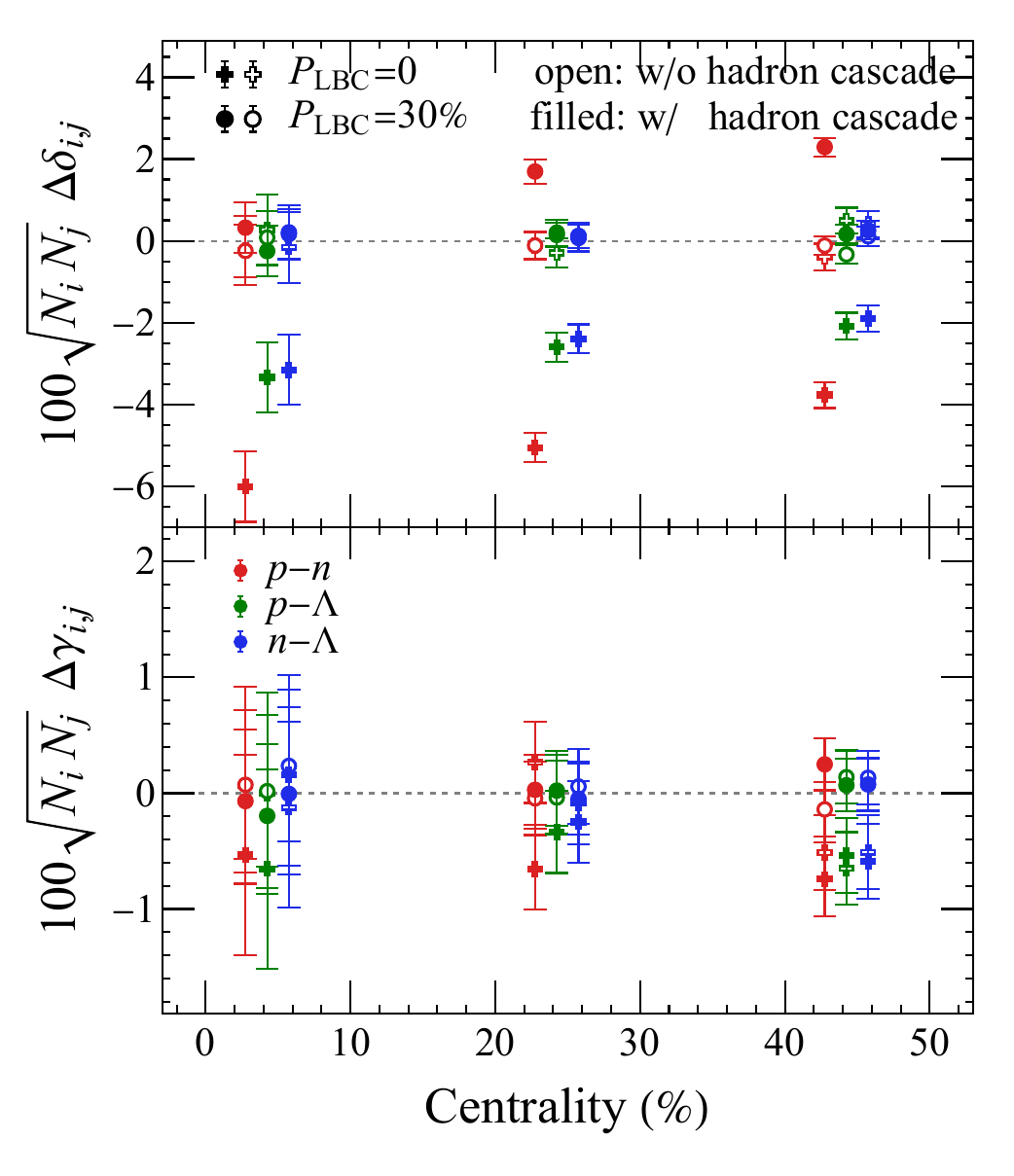}
\caption{(Color online) AVFD simulations of several baryon-baryon correlators with local baryonic charge conservation and hadron cascade enabled or disabled. The correlators on the Y-axis are scaled by corresponding particle yields.}
\label{fig:avfd}
\end{figure}

Figure~\ref{fig:avfd} shows the results for \dd and \dg obtained from AVFD+UrQMD for various baryon pairs. All correlators (on the $Y$-axis) are scaled by the number of particles.
When both LBC and hadronic cascade effects are turned off (open plus symbols), $\delta^{\mathrm{OS}}$, $\delta^{\mathrm{SS}}$, $\Delta\delta$, and $\Delta\gamma$ are all consistent with zero. Turning on LBC produces positive OS correlators as expected (open circles), while leaving SS correlators unchanged. The \dd and \dg are consequently increased, in qualitative agreement with the aforementioned BW results. 

More intriguing are the results when hadronic interactions are enabled. For the CME study, resonance decay such as $\rho^0 \to \pi^+ + \pi^-$ can generate opposite charge particle pairs and contribute significantly to electric charge correlators. In contrast, resonances decaying into a baryon-antibaryon pair are rare, however, annihilation processes ($B + \bar{B}' \to X$) remain possible. Baryons and antibaryons that are close to each other in the phase space are more likely to annihilate. This leads to negative OS correlations for all baryon species combinations, as shown in the filled symbols. Due to the small relative momentum and spatial separation between these (anti-)baryons, the annihilation effect should be enhanced when applying a narrower $\Delta\eta$ and/or a larger $\Sigma\pt$ selection. Given that significantly positive \dd and \dg are observed experimentally, we conclude that the negative contribution from annihilation alone cannot explain the data and may be masked by the stronger LBC effects.

\begin{figure}[!htb]
\includegraphics[width=\linewidth]{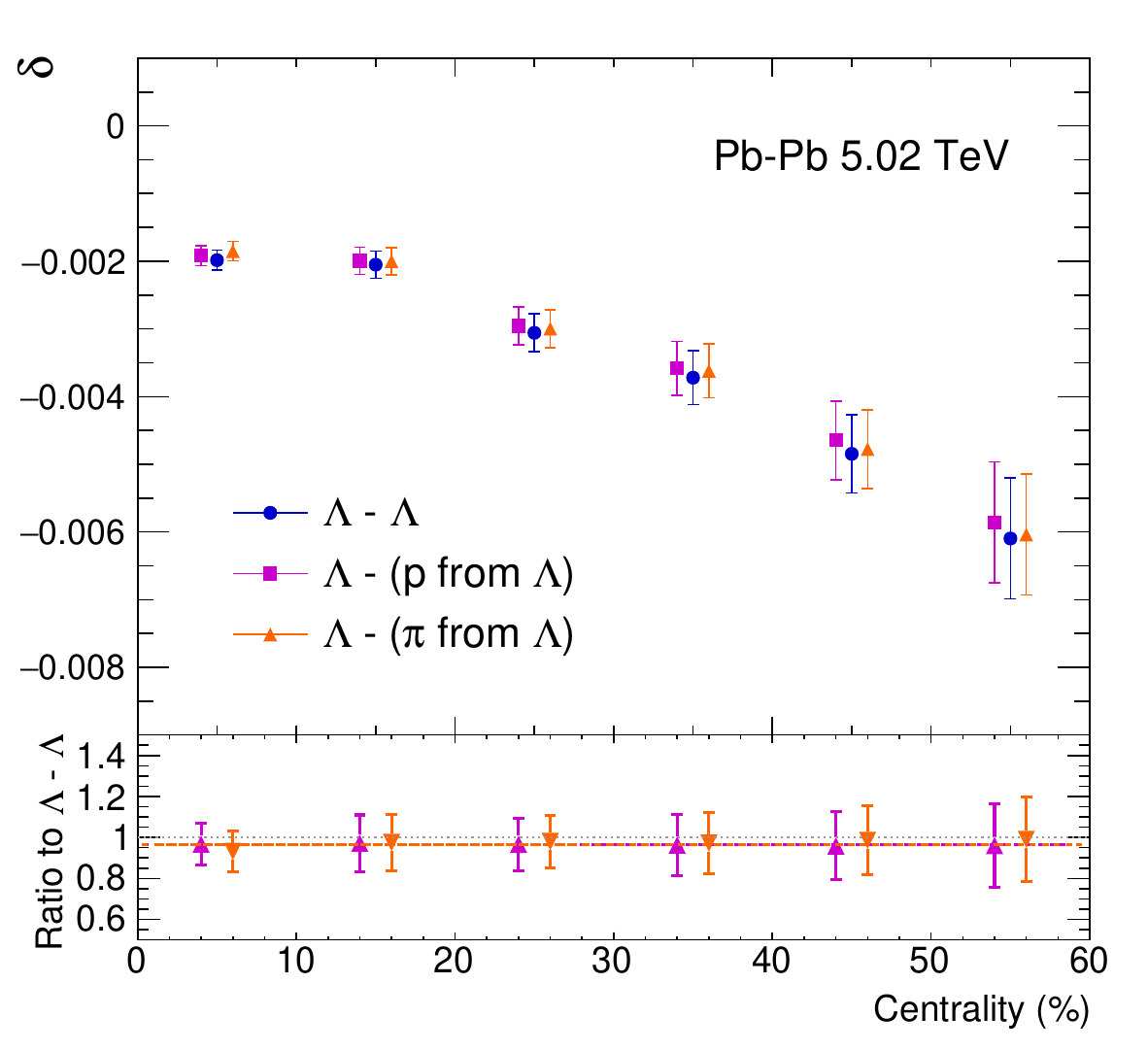}
\caption{(Color online) Feed-down contribution of secondary protons originating from $\Lambda$ decays calculated by AMPT.}
\label{fig:fig_fd}
\end{figure}

%%%%%%%%%%%%%%%%%%%%%%%%%%%%%%%%%%%%%%%%%%%%%%%%%%%%%%%%%
\subsection{Feed-down contribution from $\Lambda$ weak decays}

In experimental measurements of \lp correlations, it is possible for a $\Lambda$ to be paired with a proton originating from the decay of another $\Lambda$. The preliminary results released by ALICE have not yet incorporated feed-down correction for such contributions. Here we use the AMPT model to evaluate this effect.
Primordial $\Lambda$ are selected before the hadronic interactions and are forced to decay into $\pi$ and $p$ in their center-of-mass frame. The daughter particles were then boosted back to the laboratory frame. 
We calculate the $\delta$ correlators between two primordial $\Lambda$ particles, as well as those between one $\Lambda$ and a secondary $p$ from the decay of another $\Lambda$.
As shown in Fig.~\ref{fig:fig_fd}, the feed-down contributions are found to be nearly identical to those from $\Lambda$–$\Lambda$ correlations across all centralities, indicating that the daughters retain a significant fraction of the azimuthal information of their parent $\Lambda$. 
A constant fit over centralities yields a ratio of $\approx$97\% relative to the $\Lambda$–$\Lambda$ correlation. Therefore, in experimental analyses, the corresponding feed-down contribution can be subtracted by applying this ratio to the fraction of $p$ originating from $\Lambda$ decays~\cite{Acharya:2019}.

%%%%%%%%%%%%%%%%%%%%%%%%%%%%%%%%%%%%%%%%%%%%%%%%%%%%%%%%%
\section{Summary} \label{sum}

In this study, we employed the Blast Wave, AMPT, and AVFD+UrQMD models to investigate background contributions in the CVE measurement in Pb--Pb collisions at \snn = 5.02 TeV.
The BW model, when incorporating local baryon conservation of \lp pairs, successfully reproduces the centrality and kinematic dependences of $\Delta\delta$ and $\Delta\gamma$.
AMPT naturally captures the observed correlations without explicit tuning, indicating that they arise from initial baryon production and coalescence mechanisms.
AVFD+UrQMD simulations reveal that hadronic annihilation processes produce negative opposite-sign correlations, which are nevertheless masked by stronger LBC effects.
Furthermore, feed-down contributions from secondary protons are found to be nearly identical to those from primordial $\Lambda$–$\Lambda$ correlations.
These results highlight the dominant role of background effects, particularly LBC induced by baryon production convolved with collective expansion. We thus conclude that a robust extraction of the CVE signal will require careful subtraction of these backgrounds, and methods for effectively separating the signal from the background remain to be further explored.

%%%%%%%%%%%%%%%%%%%%%%%%%%%%%%%%%%%%%%%%%%%%%%%%%%%%%%%%%
\section*{Acknowledgments}

We are grateful to our colleagues in the ALICE and STAR Collaborations for enlightening discussions and suggestions. We thank Chen Zhong for his dedicated maintenance of the computing resources. This work is supported by the National Natural Science Foundation of China (Nos. 12322508, 12061141008, 12147101), the National Key Research and Development Program of China (Nos. 2024YFA1610802, 2024YFA1610700), and the Science and Technology Commission of Shanghai Municipality (No. 23590780100). JL is sponsored by the U.S. NSF under Grant No. PHY-2514992.

%%%%%%%%%%%%%%%%%%%%%%%%%%%%%%%%%%%%%%%%%%%%%%%%%%%%%%%%%
{}

\end{document}